\documentclass[12pt]{article}
\usepackage[dvips]{graphicx}
\newcommand \be  {\begin{equation}}  
\newcommand \bea {\begin{eqnarray} \nonumber }  
\newcommand \ee  {\end{equation}}  
\newcommand \eea {\end{eqnarray}}  
\begin{document}

\title{Stock price jumps: news and volume play a minor role}

\author{Armand Joulin$^1$, Augustin Lefevre$^{1}$ \\
Daniel Grunberg$^1$, Jean-Philippe Bouchaud$^{1}$}
%\email{bouchau@spec.saclay.cea.fr}
\maketitle

\small{\hspace{-0.7cm}
$^1$  Science \& Finance, Capital Fund Management, 6 Bd Haussmann, 75009 Paris, France.
}
 
\begin{abstract}
In order to understand the origin of stock price jumps, we cross-correlate high-frequency time series of stock 
returns with different news feeds. We find that neither idiosyncratic news nor market wide news can explain the frequency
and amplitude of price jumps. We find that the volatility patterns around jumps and around news are quite different:
jumps are followed by increased volatility, whereas news tend on average to be followed by lower volatility levels. 
The shape of the volatility relaxation is also markedly different in the two cases. Finally, we
provide direct evidence that large transaction volumes are {\it not} responsible for large price jumps.
We conjecture that most price jumps are induced by order flow fluctuations close to the point of vanishing liquidity.
\end{abstract}

Why do stock prices change? The traditional answer, within the theory of efficient markets, is that prices 
move because some new piece of information becomes available, leading to a revision of the expectations of market participants. 
If this picture was correct, and in the absence of ``noise traders'', the price should essentially be constant
between two news items, and move suddenly around the release time of the news. Noise trading should add high frequency
mean-reverting noise between news, that should not contribute to the long term volatility of the price. News 
release should be the main determinant of price volatility. There are, however, various pieces of evidence suggesting that
this picture is incorrect. Volatility is much too high to be explained only by changes in fundamentals \cite{Shiller}. 
The volatility process itself is random, with highly non-trivial clustering and long-memory properties (for reviews, see e.g. \cite{Cont,BBMZ,Book}).
Many of these properties look very similar to endogenous noise generated by complex, non-linear systems with feedback, 
such as turbulent flows \cite{Frisch,Mandel,BMD,CF,Lux}, stick balancing dynamics \cite{Stick}, etc. \cite{Intermit}. 
On liquid stocks, there is in fact little sign of high frequency mean reversion that one could attribute to noise traders \cite{subtle}. 
It rather appears that most of the 
volatility arises from trading itself, through the very impact of trades on prices.  This was the conclusion 
reached by Cutler, Poterba and Summers in an early seminal paper \cite{Poterba} (see also \cite{Fair}). More recently, the authors of 
\cite{Wyart} used high frequency data to decompose the volatility into an impact component and a news component, 
and found the former to be dominant (see also \cite{subtle,Hopman}). 
Here, we want to confirm this conclusion directly, using different news feeds synchronized with price time series. 
Our main result is indeed that most large jumps (defined more precisely below) are not related to any 
broadcasted news, even if we extend the notion of `news' to a (possibly endogenous) collective market or sector
jump. We find that the volatility pattern around jumps and around news is quite different, confirming
that these are distinct market phenomena \cite{Muzy}. We also
provide direct evidence that large transaction volumes are {\it not} responsible for large price jumps, as 
also shown in \cite{Farmer}.
We conjecture that most price jumps are in fact due to 
endogenous liquidity micro-crises \cite{Lillo}, induced by order flow fluctuations in a situation close to vanishing 
outstanding liquidity.

\begin{figure}[htbp]
    \begin{tabular}{cc}   \hspace*{-0.5cm}
    \includegraphics[width=7cm]{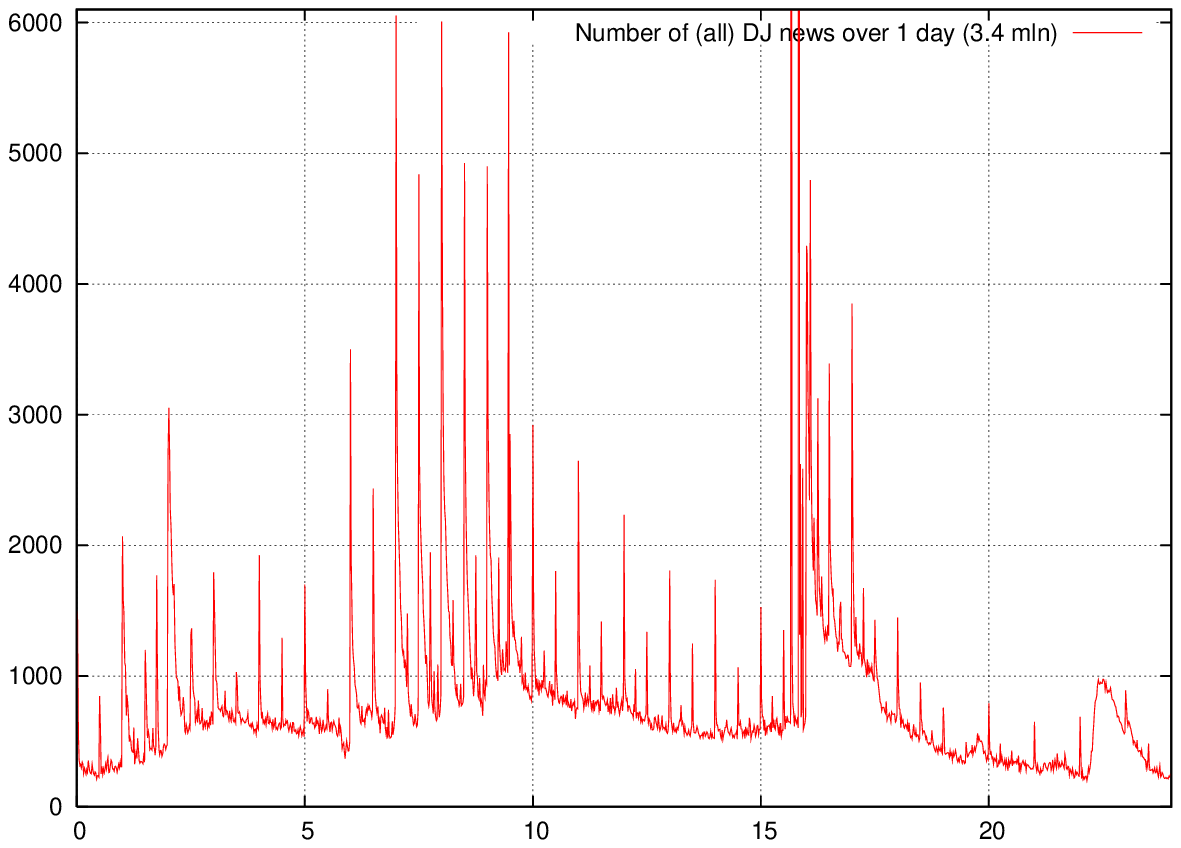} & \hspace*{-0.5cm}
    \includegraphics[width=7cm]{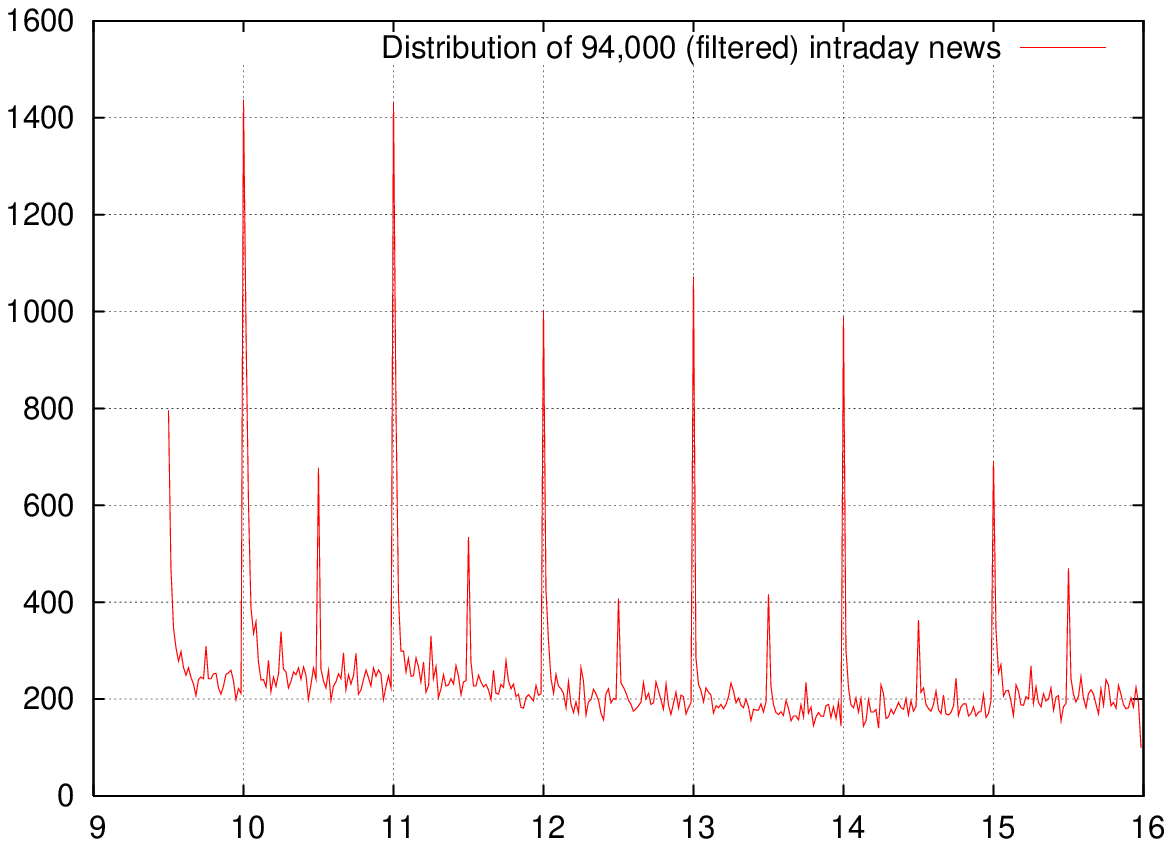} \\ \hspace*{-0.5cm}
    \includegraphics[width=7cm]{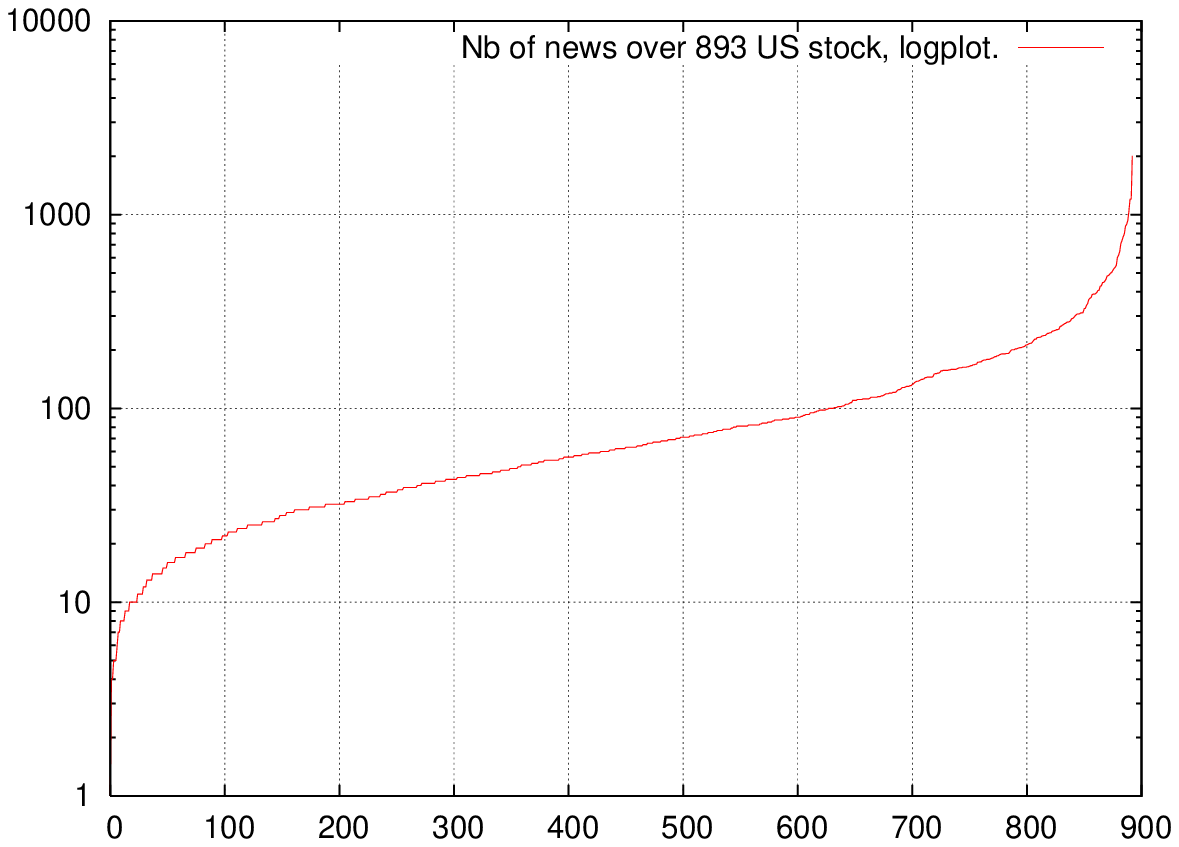} & \hspace*{-0.5cm}
    \includegraphics[width=7cm]{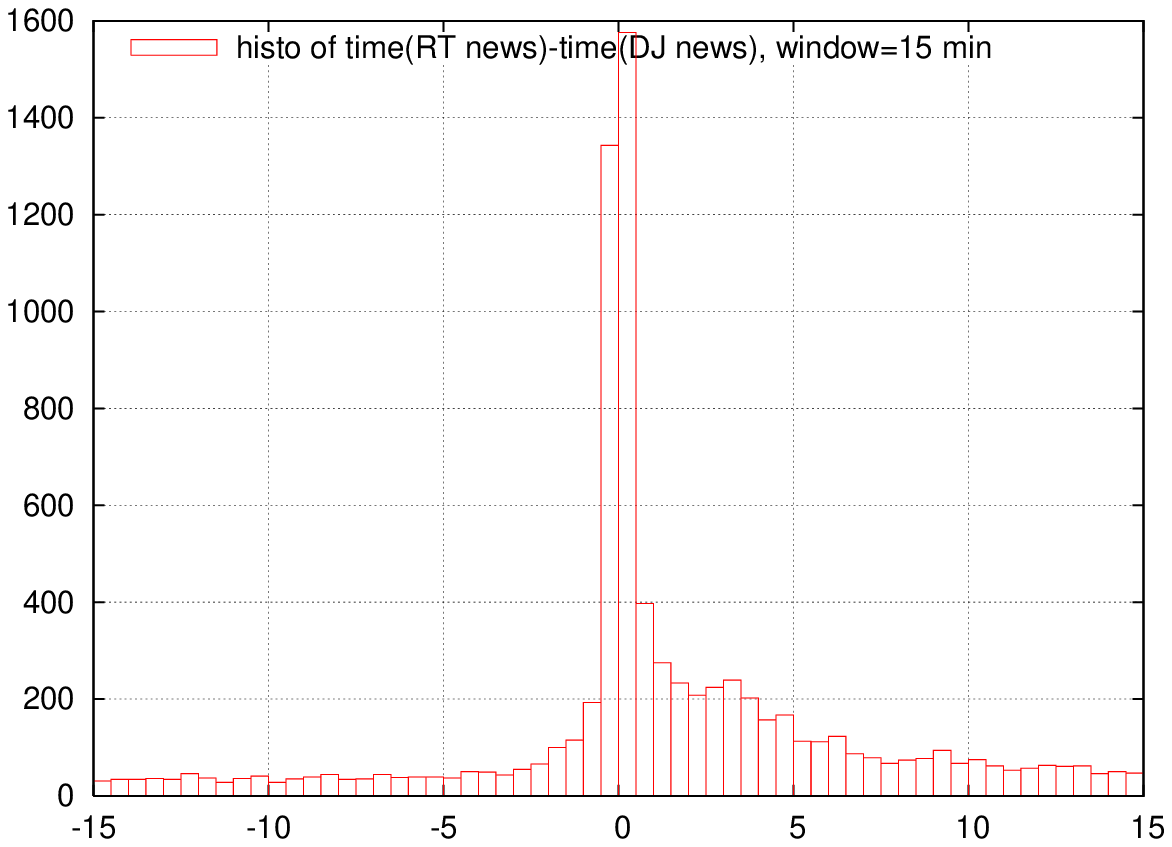}
    \end{tabular}
    \caption{Statistics on Dow Jones news for 893 US stocks over 2 years. {\bf a)} Distribution of news throughout the day (551,000 news, first in series, headline containing company's name).  
    {\bf b)} Idem for intraday, `noisy' news discarded (94,000 news).  
    {\bf c)} Distribution of number of news for each stock, throughout 893 US stocks.  
    {\bf d)} Histogram of delays between DJ news and RT news, expressed as time(RT-news)$-$time(DJ-news) in minutes.  
    For 893 US stocks and 2 years of news archive (intraday, only first news in a chain/story, noisy news filtered out), 
    we find one DJ news and one RT in the same window of 15 minutes, and require each be preceded by a silence of 30 minutes.  
    We see that DJ is quicker in 2/3 of cases (1.5 minutes ahead on avg).}
   \label{fig:news_distrib}
\end{figure}

We have studied the news feed from Dow Jones, that concerns 893 stocks from NASDAQ and NYSE, from Aug-02-2004
to Aug-31-2006, between 9:30 am to 4:00 pm. We identified a total of 93,698 relevant news items\footnote{To separate the
grain from the chaff, we took only the first news from an avalanche of news (same DJ\_id), removed generic or automated 
news (imbalances at Close, etc), and required the name of the company (among the many tickers listed) to be present 
in the headline.}, or 105 news par stock on average, or one per stock every five days. The median stock has 62 news 
(see third plot of figure~\ref{fig:news_distrib} for a distribution).  We also had the Reuters news feed at our disposal: 
32,000 news for the same 2-year period and using the same filter on the 893 stocks. Of these, only 8,000 coincide with Dow Jones news, 
i.e. there is a DJ news in a 15 min neighborhood provided both RT and DJ news are preceded by a news silence of 30 min.
From Fig.~\ref{fig:news_distrib}~d) we infer that Dow Jones is slightly faster on US stocks; 
and including the Reuters news which do not seem to have a Dow Jones counterpart increases the total number of news by 26\%. 
This does not change any of the qualitative conclusions
below. We crossed this news feed with one-minute return data files. Some of these files are however incomplete, 
so we checked all our results on a restricted subset of 163 stocks in the period Nov-28-2005 
to Jun-30-2006, for which we have a complete set of return time series. The number of news concerning these 163 
stocks is 6,443.

We first compare the occurrence of price ``jumps'' with the occurrence of news on a given company. This requires
to define jumps in a consistent, if slightly arbitrary fashion. We choose to compare the absolute size $|r(t)|$ 
of a one minute bin return to a short term (120 minutes) flat moving average $m(t)$ of the same quantity, in order to factor in slow modulations
of the average volatility. An s-jump is such that $|r(t)| > s \, m(t)$. The number of s-jumps as a function of $s$
is shown in Fig. 2; it is seen to decay as $\approx s^{-4}$, as expected from the well known approximately 
inverse cubic distribution of high frequency returns \cite{Gopi}. We note once again that this is a very broad distribution, 
meaning that the number of extreme events is in fact quite large. For example, for the already rather high
value $s=4$ and for only 166 stocks during 149 days, we found 177,674 jumps, which amounts to 1090 jumps per stock, 
or 7 to 8 jumps per stock per day! 
A threshold of $s=8$ decreases this number by a factor $\approx 10$, amounting to one jump every one day and 
a half per stock. On the same period and set of stocks we found 6443 news, or one news every 3 days for each stock.
These numbers already suggest that a very large proportion of shocks (say, 4-jumps) cannot be attributed to idiosyncratic
news (i.e. a news item containing the ticker of a given stock). The possibility of more collective news affecting
macroeconomic variables will be discussed below.

\begin{figure}[htbp]
\centering
\resizebox{10cm}{!}{\includegraphics[trim = 0 -40 0 -60]{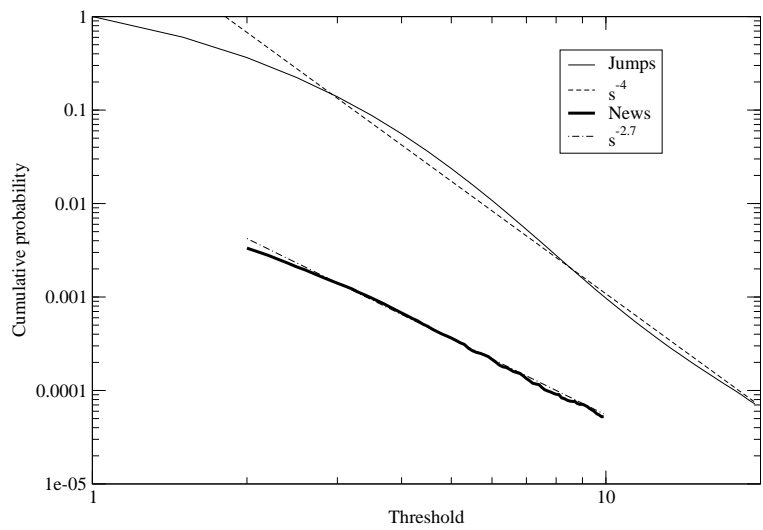}}
\caption{Cumulative distribution of s-jumps as a function of $s$. The scale is log-log. The distribution decays as $\approx s^{-4}$. We also show the 
number of s-jumps associated with news. Interestingly, this distribution also decays as a power-law but with a smaller exponent, 
$\approx s^{-2.7}$.}
\label{jump_distrib}
\end{figure}

\begin{figure}[htbp]
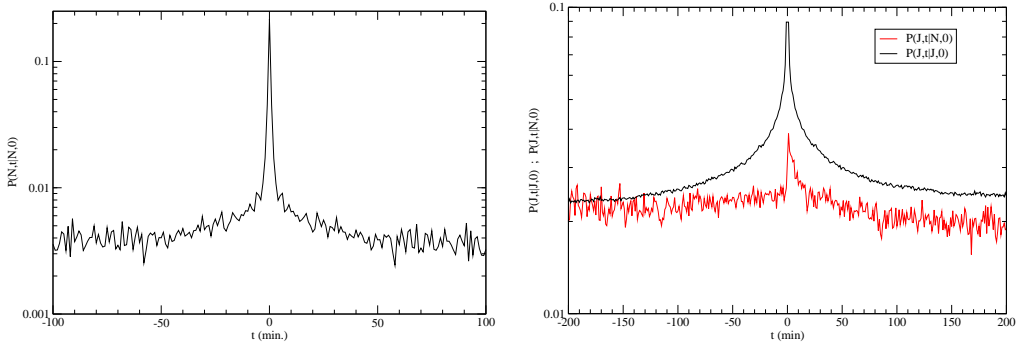

    \begin{tabular}{cc}   
    \includegraphics[width=6.5cm]{PNtN0.eps} &
    \includegraphics[width=6.5cm]{PJtJ0.eps} 
    \end{tabular}
    \caption{Clustering of news and jumps, measured by conditional probabilities 
    {\bf a)} Prob(news at time t $|$ news at time 0), 200 minutes before and after $t=0$, 
    ignoring overnight news. 
    {\bf b)} Prob(jump at time t $|$ jump at time 0) and Prob(jump at time t $|$ news at time 0). Jumps here correspond to $s=4$.
    Note that the probability to observe a jump after a news is actually decreased at intermediate times. 
    For all densities, we divided each contribution by the average daily distribution of news or jumps, such as to remove
    intra-day seasonalities.}
   \label{fig:news_clustering}
\end{figure}

In order to be more quantitative, we show in figure~\ref{fig:news_clustering} a-b the conditional probability to observe a news (resp. a jump) 
at time $t$ knowing that there was a news (resp. a jump) at time $t=0$. One notices a substantial 
level of clustering in both cases. In figure~\ref{fig:news_clustering}~b, we also show the probability
to observe a jump at time $t$ conditioned to a news at time $t=0$.  Here, we see only a very small, but 
significant, asymmetric increase of the probability to see a jump induced by a news.  This probability is only
increased by a factor $\approx 2$ in the two minutes following the news compared to the average background. So 
not only most jumps are not induced by news: most news do not induce any real jump at all!  This decoupling means that most news are 
either expected (through rumors and leakage) or deemed insignificant by the market: genuine surprising intraday news are rare. Note that 
companies indeed tend to publish big surprises -- like earnings -- in overnight. Even earning forecasts by analysts are not taken very
seriously by the market (see \cite{Guedj} for a discussion of this point).

\begin{figure}[htbp]
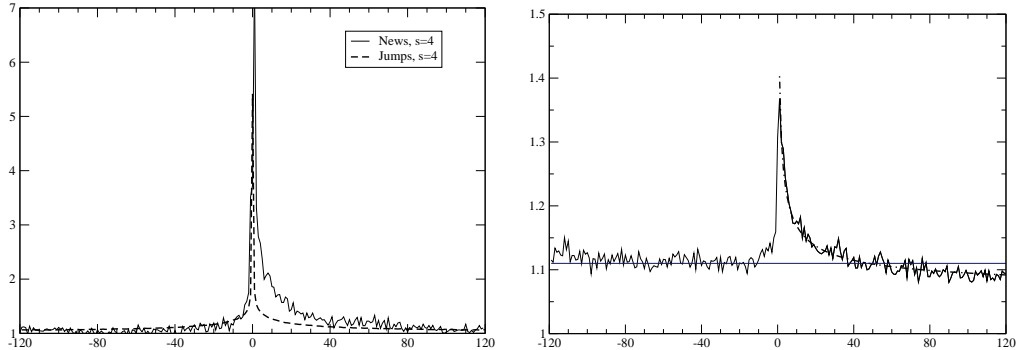

    \begin{tabular}{cc}  
    \includegraphics[width=6.5cm]{News-Jumps4.eps} & 
    \includegraphics[width=6.5cm]{News.eps}

    \end{tabular}
    \caption{{\bf a)} Volatility 120 minutes before and after the $s=4$ news and $s=4$ jumps.
    We divided each contribution by the U-shaped daily volatility pattern. Note that the news peak is 
    much broader, but diverges faster for $t \to 0^+$ (see Fig.~\ref{fig:vol2}).
    {\bf b)} Volatility 120 minutes before and after intraday news, averaged over {\it all} significant news. 
    Note that the long time limit falls {\it below} the pre-news volatility.}
   \label{fig:vol}
\end{figure}

\begin{figure}[htbp]
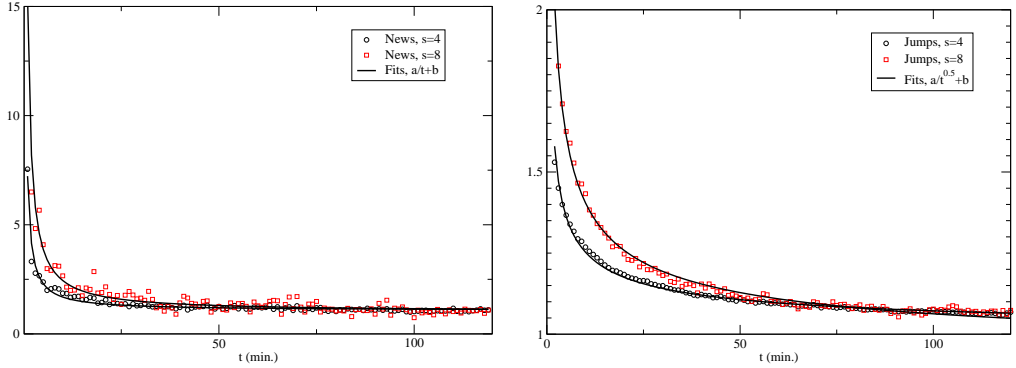

    \begin{tabular}{cc}  
    \includegraphics[width=6.5cm]{Newsrelax.eps}&
    \includegraphics[width=6.5cm]{Jumprelax.eps}

    \end{tabular}
    \caption{{\bf a)} Relaxation of the volatility after $s=4$ and $s=8$ news jumps, and power law fit with an exponent $\beta=1$.
    {\bf b)} Relaxation of the volatility after $s=4$ and $s=8$ jumps, and power law fit with an exponent $\beta=1/2$. Note the difference
    in the y-scale between the two curves (the first point, at $t=0$, is suppressed in both cases).}
   \label{fig:vol2}
\end{figure}

Another clear-cut difference between jumps and news is the volatility pattern around the two types of events. In 
Fig.~\ref{fig:vol} a), we show the average absolute one-minute return $\sigma(t)=\langle |r(t)|\rangle$ conditioned to the presence of an
$s=4$ jump at $t=0$, induced by a news or not. We can in fact determine the size distribution of news induced jumps. 
We find again a power law $s^{-\mu}$, but with an exponent $\mu \approx 2.7$ (see Fig. 1), clearly different from the value $\mu = 4$ mentioned above
for jumps. 
The volatility pattern in the case of news is much wider than the rather narrow peak 
corresponding to endogenous jumps. In both cases, we find (Figure~\ref{fig:vol2}) that the relaxation of the excess-volatility follows a power-law
in time $\sigma(t)-\sigma(\infty) \propto t^{-\beta}$ (see also \cite{Kertecz,Mantegna}). The exponent of the decay is, however, markedly different in the two cases: for news 
jumps, we find $\beta \approx 1$, whereas for endogenous jumps one has $\beta \approx 1/2$. Our results are compatible with those of 
\cite{Kertecz}, who find $\beta \approx 0.35$. The difference between endogenous and endogenous volatility relaxation has also
been noted in \cite{Muzy}, but on a very restricted set of news events. We found the value of $\beta$ to be identical, within error bars, for
the two threshold values $s=4$ and $s=8$ (see Fig.~\ref{fig:vol2}), although a three parameter fit is compatible with  
$\beta(s=8) > \beta(s=4)$ for jumps, as predicted by the multifractal random walk model \cite{Muzy} and the multiscale feedback
model of \cite{Lisa}.

If we now average the volatility pattern around {\it all} news, we
find a rather modest peak height (the increase is only 30-40\%
of the baseline level), confirming the result of figure~\ref{fig:news_clustering}~b: news are often of no real importance and 
only mildly affect the price of a stock. The volatility decays with an exponent $\beta \approx 1/2$ towards a baseline value 
that is on average {\it lower} than before the news, as if news release had, on average, a sedating effect on markets. A similar effect 
can be noted in Fig. \ref{fig:news_clustering}~b. This is however not true of jumps or of `strong' news (corresponding to $s \geq 4$ jumps): 
both endogenous jumps and unexpected news shake the market and induce extra uncertainty, that -- as noted above -- decays differently for jumps and for news. 

The conclusion of this first part is that jumps and news appear to be, in general, distinct events, both 
concerning the amplitude of the volatility increase, and the shape of the volatility profile around these 
events. Jumps seem to occur for no identifiable reasons, and as a consequence spook the market; this perturbation 
is slow to decay (as $1/\sqrt{t}$). Although counter-intuitive 
at first, news seem (on average) to reduce any measure of uncertainty, since some previously unknown information becomes 
available. However, some news are really significant and also lead to an increased future volatility. 

The above analysis focused on idiosyncratic company news only. However, some important macroeconomic news 
might affect the market (or a sector) as a whole and induce important price jumps that we fail to identify
in our company news data set. Of course, the identification of possibly relevant macroeconomic news is 
at first sight much more difficult, since this would entail contextual word recognition. A short cut is to
consider that any important macroeconomic news concerning e.g. interest rates, oil prices, inflation, etc. 
should induce a `collective jump' in the market (see below for a precise definition). From the
work of Cutler et al. \cite{Poterba}, we already know that some of the market jumps are not explained
by clearly identified macroeconomic events. But we could in fact {\it define} a market jump as being news, to
which individual stocks might be sensitive to. 

There are different possible definitions of collective jumps. One is based on the correlation matrix of 
jump occurrences, $c_{ij}=T^{-1}\sum_t \theta_i^t \theta_j^t - p_i p_j$, where $\theta_i^t$ is one if 
stock $i$ makes an s-jump in the one minute bin $t$ and zero otherwise, $p_i$ is the average of $\theta_i^t$ (i.e.
the jump probability $p_i=T^{-1}\sum_t \theta_i^t$) and $T$ is the number of bins. Most eigenvalues of $c$ are seen to lie within the Marcenko-Pastur 
noise band \cite{MP,Cracow}, 
but a few stand out, in particular the `market' eigenvalue with a eigenvector $v_i^1$ close to uniform across all 
stocks: $v_i^1 \approx N^{-1/2}$, $\forall i$. A market jump can be defined such that the indicator $\chi^t= N^{1/2} 
\sum_i \theta_i^t v_i^1$ is larger than a certain threshold $ s'$. For example $s'=0.1$ means that 
roughly $10 \%$ of the stocks must jump to qualify as a market jump. Quite interestingly, we found a new 
`stylized fact' concerning index jumps: the cumulative distribution of $\chi$ decays as a power-law $\chi^{-\nu}$ with exponent $\nu \approx 1.5$. 
In other words, the number of stocks involved in a market slide is very broadly distributed.
Another definition of collective jumps is based on a standard sector classification to define a sector index; 
a sector jump is such that this index exceeds a certain threshold. 

Focussing first on market jumps, and for a threshold $s'=0.1$ that defines rather loose collective jumps, we 
find a total of $900$ jumps, but only $13 \%$ of all individual jumps with $s=4$ can be explained by these jumps, hardly more than
the $10 \%$ expected from the very definition of these jumps. Extending the period around collective jumps to
a five minute interval, one can increase this number to $20 \%$, but this increase is the one expected from
the unconditional probability of jumps, integrated over 5 minutes. If we include further meaningful sectors, we increase the above $13 \%$ 
of explained jumps up to $21 \%$. Increasing the threshold to $s'=0.3$ further decreases these numbers to less than $5 \%$ even 
with a five minute interval. Changing 
the definition of collective jumps does not change the above picture.

\begin{figure}[htbp]
\centering
\resizebox{10cm}{!}{\includegraphics[trim = 0 -40 0 -60]{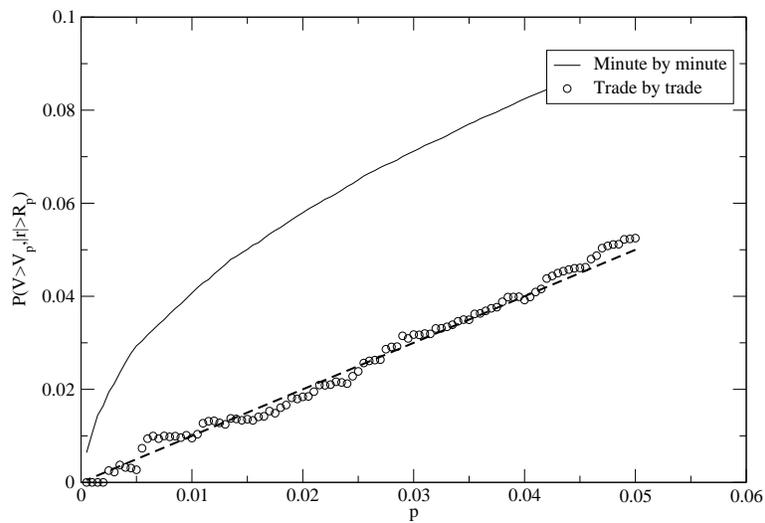}}
\caption{Tail correlations between returns and volumes, both at the trade by trade level (for 196 NASDAQ stocks between
June 12,2006 and August 3, 2006) and at the one-minute bin level (averaged over all stocks of the above data set). 
Clearly, in both cases, ${\cal C}(p \to 0) \to 0$.}
\label{ret_vol}
\end{figure}

The conclusion here is that even if one associates every collective jump to real macroeconomic news (which we 
know not be the case), a significant fraction of individual stock jumps are still unexplained. It therefore 
seems clear that the explanation of these jumps must lie elsewhere. One idea is that private information
could play a role and trigger these jumps. But if an investor really had valuable private information he would
trade as to reveal as little as possible of this information. This is best exemplified by Kyle's model of 
informed trading \cite{Kyle}, where the trades of the insider are perfectly hidden in the uninformed flow. 
Nonetheless, it was recently 
claimed that large price jumps are due to large incoming volumes that destabilize the order book and lead to
large swings \cite{Gabaix}. This point of view was rather convincingly challenged in \cite{morethanvolume}. Here, we confirm
explicitly that
large price jumps are in fact {\it not} induced by large transaction volumes. We do this by studying tail correlations
between absolute value of returns and volume. Define $R_p$ and $V_p$ as the p-quantiles of absolute returns and 
transaction volumes, i.e.:
\be
P(|r| > R_p) = p \qquad P(V > V_p)=p.
\ee
Now, define the tail correlations as:
\be
{\cal C}(p)= P(|r| > R_p | V > V_p) = P(V > V_p ||r| > R_p).
\ee
If $|r|$ and $V$ are independent, then clearly ${\cal C}(p)=p$.
Since both $|r|$ and $V$ are power-law distributed, a relation of the type $|r| = \lambda V^\alpha + \epsilon$, as postulated
in \cite{Gabaix}, leads to a finite limit for ${\cal C}(p)$ when $p \to 0$ whenever the residual $\epsilon$ is asymptotically sub-dominant
(see e.g. \cite{Book}, Ch. 11). 
This would be true even if only a finite fraction
of the events followed such a relation. If ${\cal C}(p \to 0) \to 0$, on the other hand, this would mean that extreme
returns and extreme volumes are asymptotically independent. We have studied ${\cal C}(p)$ both at the trade by trade level and
for one-minute aggregate returns and volumes. The results are given in Fig. \ref{ret_vol}. Clearly, at the trade by trade level, 
there is {\it no} extreme correlations at all since ${\cal C}(p) \approx p$ to a good approximation. For one minute bins, there are some 
significant correlations (${\cal C}(p) >  p$) that are anyway expected since large price changes tend to be followed by large volumes of 
activity. But still, it is very clear that ${\cal C}(p)$ tends to zero when $p \to 0$, meaning that large jumps are {\it not} induced by large trading
volumes. 

So what is left to explain the seemingly spontaneous large price jumps? We believe that the explanation comes from
the fact that markets, even when they are `liquid', operate in a regime of {\it vanishing liquidity}, and therefore
are in a {\it self-organized critical state} \cite{Bak}.
On electronic markets,  the total volume available in the order book is, at any instant of time, 
a tiny fraction of the stock capitalisation, say $10^{-5}-10^{-4}$ (see e.g. \cite{Wyart}).\footnote{This 
explain why liquidity takers have to cut their orders in small pieces, thereby creating long range correlations
in order flow \cite{subtle,Farmer}.} Liquidity providers take the risk of 
being ``picked off'', i.e. selling just before a big upwards move or vice versa, and therefore place limit 
orders quite cautiously, and tend to cancel these orders as soon as uncertainty signals appear. Such signals
may simply be due to natural fluctuations in the order flow, which may lead, in some cases, to a catastrophic decay in
liquidity, and therefore price jumps. There is indeed evidence that large price jumps are due to local 
liquidity dry outs. It would be very interesting to find a simple order flow model that captures these 
spontaneous critical liquidity fluctuations. Such a model would usefully shed light on the controversial nature
of price jumps and help understand their origin.

\vskip 1cm

{\bf Acknowledgements} We thank F. Bonnevay, Z. Eisler, J. D. Farmer, J. Kockelkoren, L. Laloux, M. Potters and V. Vargas for 
inspiring discussions and help.

\end{document}